\newcommand{\rem}[1]{}
\documentclass{amsart}
\usepackage{amsfonts,amssymb,amsmath,amsthm,mathrsfs}
\usepackage{url}
\usepackage[dvips]{epsfig}
\newtheorem{thrm}{Theorem}[section]

\newtheorem{prop}[thrm]{Proposition}
\newtheorem{cor}[thrm]{Corollary}
\newtheorem{remark}[thrm]{Remark}
\theoremstyle{definition}

\begin{document}
\author[L.~G.~Molinari, C.~A.~Mantica]
{Luca~Guido~Molinari and Carlo~Alberto~Mantica }
\address{L.~G.~Molinari (corresponding author): Physics Department,
Universit\`a degli Studi di Milano and I.N.F.N. sez. Milano,
Via Celoria 16, 20133 Milano, Italy -- C.~A.~Mantica: I.I.S. Lagrange, Via L. Modignani 65, 
20161, Milano, Italy, and I.N.F.N. sez. Milano.}
\email{Luca.Molinari@unimi.it, Carlo.Mantica@mi.infn.it}
\begin{abstract}
We prove that, in a space-time of dimension $n>3$ with a velocity field that is shear-free,
vorticity-free and acceleration-free, the covariant divergence of the Weyl tensor is zero if the contraction of the 
Weyl tensor with the velocity is zero. The other way, if the covariant divergence of the Weyl tensor is
zero, then the contraction of the Weyl tensor with the velocity has recurrent geodesic derivative.
This partly extends a property found in Generalised Robertson-Walker spacetimes, where the
velocity is also eigenvector of the Ricci tensor.
Despite the simplicity of the statement, the proof is involved. \\
As a product of the same calculation, we introduce a curvature tensor with an interesting
recurrence property.
\end{abstract}
\title[A simple property of the Weyl tensor ]
{A simple property of the Weyl tensor\\
for a shear, vorticity and acceleration-free 
velocity field}
\maketitle

\section{Introduction}
A shear-free, vorticity-free and acceleration-free velocity field $u_k$, has covariant derivative  
\begin{align}
\nabla_i u_j = \varphi \, (g_{ij} + u_i u_j) \label{torseforming}
\end{align}
where $\varphi $ is a scalar field, and $u_k u^k=-1$. For such a vector field we prove the following results
for the Weyl tensor, in space-time dimension $n>3$:
\begin{thrm}\label{thrm_iff}
\begin{align}
i)\quad  u_m C_{jkl}{}^m=0 \;&\Longrightarrow\; \nabla_m C_{jkl}{}^m=0 \\
ii) \quad \nabla_m C_{jkl}{}^m=0 \;&\Longrightarrow\;  u^p\nabla_p (u_m C_{jkl}{}^m)= - \varphi (n-1) u_m C_{jkl}{}^m
\end{align}
\end{thrm}
Next, we introduce the following tensor, where $E_{kl}=u^ju^m C_{jklm}$ is the electric part of the Weyl tensor:
\begin{align}
\Gamma_{iklm} = C_{iklm} - \frac{n-2}{n-3} (u_i u_mE_{kl}-u_k u_mE_{il}  -u_iu_l E_{km}+u_k u_l E_{im} ) \label{Gammatensor}
\\ 
-\frac{1}{n-3}( g_{im} E_{kl}  - g_{km} E_{il} - g_{il} E_{km} + g_{kl} E_{im}  ) \nonumber
\end{align}
\begin{thrm}\label{th_gamma}
$\Gamma_{jklm}$ is a generalised curvature tensor, it is totally trace-less and:
\begin{gather}
u^m \Gamma_{jklm} =0\\
u^p\nabla_p \Gamma_{jklm} = - 2\varphi  \Gamma_{jklm} 
\end{gather}
\end{thrm}
\noindent
The tensor is zero in $n=4$.

The proofs make use of various properties of ``twisted'' space-times,
that were introduced by B.~Y.~Chen \cite{Chen79} as a generalisation of warped space-times: 
\begin{align}
ds^2 = -dt^2 + f^2(\vec x,t) g^*_{\mu\nu} (\vec x) dx^\mu dx^\nu \label{twisted}
\end{align}
$f>0$ is the scale factor and $g^*_{\mu\nu}$ is the metric tensor of a Riemannian sub-manifold of dimension $n-1$. If $f$ only depends on time, the metric is warped and the space-time is a Generalized Robertson-Walker (GRW) space-time \cite{AliRomSan95_a,Chen14,survey}. 
Chen \cite{Chen17} and the authors \cite{ManMolGRG} gave covariant characterisations of twisted space-times; the latter reads: {\em a space-time is twisted if and only if there exists a time-like unit vector field $u^i$ with the property \eqref{torseforming}.}\\  
The space-time is GRW if $u^i$ is also eigenvector of the Ricci tensor \cite{survey}; it is RW with the further condition that the Weyl tensor is zero, $C_{jklm}=0$.

The next two short sections collect useful results on twisted space-times, and about the Weyl tensor in $n=4$. 

\section{Twisted space-times}
We summarise some results on twisted space-times, taken from ref. \cite{ManMolGRG}:\\
i) the vector field $u_j$ is Weyl compatible (see \cite{ManMol2014} for a general presentation):
\begin{align}
(u_iC_{jklm}+u_j C_{kilm}+u_kC_{ijlm}) u^m =0. \label{WeylComp}
\end{align}
This classifies the Weyl tensor as purely electric with respect to $u_j$ \cite{HOW}.\\
A contraction gives the useful property:
\begin{align}
C_{jklm}u^m =  u_k E_{jl} -u_j E_{kl} \label{CCC}
\end{align}
where $E_{jk} = C_{ijkl}u^iu^l$. It follows that $C_{jklm}u^m=0$ if and only if $E_{ij}=0$.\\
ii) the Ricci tensor has the general form 
\begin{align}
R_{jk} =  \frac{R-n\xi}{n-1} \, u_j u_k  + \frac{R-\xi}{n-1} g_{jk}  + (n-2)(u_jv_k+u_k v_j - E_{jk})  \label{eq_Ricci}
\end{align}
where  $R=R^k{}_k$, $\xi=(n-1) (u^p\nabla_p \varphi +\varphi^2)$,  and $v^k =(g^{km}+u^ku^m)\nabla_m \varphi $ is a space-like vector.\\ 
iii) A twisted space-time is a GRW space-time if and only if $v_j=0$.

\section{The Weyl tensor in four-dimensional space-times}
The following algebraic identity by Lovelock holds in $n=4$ (\cite{Lovelock-Rund}, ex. 4.9):
\begin{align}\label{eq_LR}
0&= g_{ar} C_{bcst} + g_{br} C_{cast}+ g_{cr} C_{abst} \\
&+g_{at} C_{bcrs} + g_{bt} C_{cars}+ g_{ct} C_{abrs} \nonumber \\
&+g_{as} C_{bctr} + g_{bs} C_{catr}+ g_{cs} C_{abtr} \nonumber
\end{align}
It implies that
$C_{abcr} C^{abcs} = \tfrac{1}{4} \delta_r{}^s C^2 $, where $C^2 = C_{abcd}C^{abcd}$.\\
The contraction of \eqref{eq_LR} with $u^cu^r$, where $u^j$ is any time-like unit vector, gives the Weyl tensor in terms of its contractions $u^d C_{abcd}$ and 
$E_{ad} = u^bu^c C_{abcd}$:
\begin{align}
C_{abcd}= - u^m (u_a C_{mbcd} + u_b C_{amcd} + u_c  C_{abmd}+ u_d  C_{abcm} ) \label{CABCD}\\
+g_{ad} E_{bc} - g_{bd} E_{ac} -g_{ac} E_{bd} + g_{bc} E_{ad}\nonumber
\end{align}
\begin{prop}\label{prop_Weylcomp_n4}
If $u^m$ is Weyl compatible, \eqref{WeylComp}, in $n=4$ the Weyl tensor is wholly given
by its electric component:
\begin{align}
 C_{abcd} =  2(u_au_d E_{bc} -u_au_c E_{bd}+u_bu_c E_{ad}- u_bu_d E_{ac})\\
 + g_{ad} E_{bc}- g_{ac} E_{bd}  + g_{bc} E_{ad} - g_{bd}E_{ac} \nonumber 
 \end{align}
 and $C^2 = 8 \, E^2$, where $E^2=E_{ab}E^{ab}$.
\begin{proof}
The property \eqref{CCC} is used to simplify \eqref{CABCD}.
Contraction with $u^iu^j$ of the identity $ \frac{1}{4}C^2g_{ij} = C_{iabc}C_j{}^{abc}  $ and \eqref{CCC} give:
$ -\tfrac{1}{4} C^2= (u^i C_{iabc})(u_j C^{jabc})  =(u_bE_{ca}- u_c E_{ba})(u^bE^{ca}- u^c E^{ba})$.
Since $E_{ca}u^c=0$, the result is $-E_{ca}E^{ca}-E_{ba}E^{ba}=-2E^2$.
\end{proof}
\end{prop}
\begin{cor}
In a twisted space-time in $n=4$,  $C_{abcd}=0$ if and only if $E_{ab}=0$.
\end{cor}

\section{The main results}
In $n>3$ the second Bianchi identity for the Riemann tensor translates to an identity for the Weyl tensor \cite{Adati}:
\begin{align}\label{Adati}
\nabla_i C_{jklm} + \nabla_j C_{kilm} + \nabla_k C_{ijlm}
= \tfrac{1}{n-3} \nabla_p
(g_{jm}  C_{kil}{}^p + g_{km} C_{ijl}{}^p \nonumber \\
+ g_{im}  C_{jkl}{}^p
+ g_{kl}  C_{jim}{}^p + g_{il} C_{kjm}{}^p + g_{jl}  C_{ikm}{}^p ).
\end{align}
As a consequence of \eqref{Adati}, as shown in the Appendix, we obtain
the intermediate result:
\begin{prop}
In a twisted space-time the divergence of the Weyl tensor is:
\begin{align}\label{divCtwisted}
 \nabla_p C_{ikm}{}^p &=(n-3)(\nabla_i  E_{km} - \nabla_k E_{im})\\
& \quad + (n-2) [ u^p\nabla_p (u_i E_{km}-u_k E_{im} ) +2 \varphi (u_iE_{km} - u_k E_{im})]
 \nonumber\\
& \quad + (2u_k u_m +g_{km}) \nabla_p E_i{}^p - (2u_i u_m +g_{im}) \nabla_p E_k{}^p. \nonumber
\end{align}
\end{prop}
\begin{cor} In a twisted space-time, if  $\nabla^p C_{jklp}=0$ then
\begin{align}
 \nabla_p E^{pk}=0  \quad \text{and}\quad
 u^p\nabla_p E_{km}= -\varphi (n-1) E_{km}  \label{id_cor}
\end{align}
\begin{proof}
Note the identity: $u^m \nabla_p C_{jkm}{}^p= \nabla_p (u^m C_{jkm}{}^p) =
\nabla_p(u_j E_k{}^p-u_kE_j{}^p)=u_j \nabla_p E_k{}^p-u_k\nabla_p E_j{}^p$. Then:
$u^k u^m \nabla_p C_{jkm}{}^p= \nabla_p E_j{}^p$.\\
Another identity is:
$u^j \nabla_p C_{jkm}{}^p= \nabla_p (u^j C_{jkm}{}^p) -\varphi E_{km}= \nabla_p(u_m E^p{}_k-u^pE_{mk})-\varphi E_{km}
= u_m \nabla_p E^p{}_k -\varphi (n-1) E_{km} -u^p\nabla_p E_{km}$.\\
Together, the two identities imply the statements.
\end{proof}
\end{cor}
\noindent
Now, we are able to extend to twisted space-times a property of GRW space-times (Theorem 3.4, \cite{ManMolJMP}):

{\bf Theorem \ref{thrm_iff}}: In a twisted space-time of dimension $n>3$: 
\begin{align}
i)\quad  u_m C_{jkl}{}^m=0 \;&\Longrightarrow\; \nabla_m C_{jkl}{}^m=0 \label{equiv1}\\
ii) \quad \nabla_m C_{jkl}{}^m=0 \;&\Longrightarrow\;  u^p\nabla_p (u_m C_{jkl}{}^m)= - \varphi (n-1) u_m C_{jkl}{}^m \label{equiv2}
\end{align}
\begin{proof}
If $u^m C_{jklm}=0$ then $E_{kl}=0$ and $\nabla_m C_{jkl}{}^m=0$ follows from \eqref{divCtwisted}.\\
Consider the identity \eqref{CCC} for the Weyl tensor. Then:
$$ u^p\nabla_p(C_{jklm}u^m )= u_k u^p\nabla_p E_{jl}-u_j u^p\nabla_p E_{kl} $$
If $\nabla^m C_{jklm}=0$ eq.\eqref{id_cor} holds, and \eqref{equiv2} is proven.
\end{proof}
\begin{remark}\quad\\
- In the special case of generalised Robertson-Walker space-times the stronger statement 
$ u_m C_{jkl}{}^m =0 \Longleftrightarrow \nabla_m C_{jkl}{}^m  =0 $
holds (theorem 3.4 of ref.\cite{ManMolJMP}).\\
- If $n>3$ and $\nabla_m C_{jkl}{}^m =0$, the identities \eqref{id_cor} simplify eq.\eqref{divCtwisted} as follows: 
$$ \nabla_i  E_{km} - \nabla_k E_{im} = (n-2) \varphi (u_iE_{km} - u_k E_{im})  $$
Recurrence conditions of this kind are studied in theorem 6 of ref.\cite{ManSuh}.
\end{remark}
The final result \eqref{Gamma123} in the Appendix, suggests the introduction of the new tensor \eqref{Gammatensor}, that combines the Weyl tensor with the generalized curvature tensors obtained as 
Kulkarni-Nomizu products of $E_{ij}$ with $u_iu_j$ or $g_{ij}$.\\
It has the symmetries of the Weyl tensor for exchange and contraction of indices, 
as well as the first Bianchi identity (it is a generalized curvature tensor). Moreover it is traceless, $\Gamma_{mbc}{}^m=0$,
and any contraction with $u$ is zero.\\ 
The associated scalar $\Gamma^2 = \Gamma_{abcd}\Gamma^{abcd}$ is evaluated:
\begin{align}
 \Gamma^2 = C^2 - 4\frac{n-2}{n-3} \, E^2  \label{gammasq} 
 \end{align}
By Prop. \ref{prop_Weylcomp_n4} this tensor is identically zero in $n=4$.\\ 
In dimension $n>4$, {\bf Theorem \ref{th_gamma} }
is basically the result \eqref{Gamma123} of the long calculation in the Appendix.
%
\begin{remark} 
The property $\Gamma_{abcd} u^d=0$ means that in the frame \eqref{twisted}, where $u^0=1$ and space components $u^\mu$
vanish, the components $\Gamma_{abcd} $ where at least one index is time, are zero. Therefore, $\Gamma^2>0$ 
in $n>4$ and, for the same reason, $E^2\ge 0$. 
We conclude that the Weyl scalar is positive:
\begin{align}
 C^2 = 4\frac{n-2}{n-3} \, E^2 +  \Gamma^2 \ge 0
\end{align}
\end{remark}
\section*{Appendix}
\begin{prop}
In a twisted space the following identities hold among the Weyl tensor and the contracted Weyl tensor:
\begin{align}\label{2divCtwisted}
& \nabla_p C_{ikm}{}^p=(n-3)(\nabla_i  E_{km} - \nabla_k E_{im})\\
& \quad + (n-2) [ u^p\nabla_p (u_i E_{km}-u_k E_{im} ) +2 \varphi (u_iE_{km} - u_k E_{im})]
 \nonumber\\
& \quad + (2u_k u_m +g_{km}) \nabla_p E_i{}^p - (2u_i u_m +g_{im}) \nabla_p E_k{}^p \nonumber
\end{align}
\begin{align}\label{Gamma123}
& (n-3)( u^p\nabla_p C_{iklm} + 2\varphi C_{iklm} ) \\
& =   (n-2) [ u^p\nabla_p (u_i u_mE_{kl}-u_k u_mE_{il}  -u_iu_l E_{km}+u_k u_l E_{im} ) \nonumber \\
& \qquad\qquad +2\varphi  (u_iu_mE_{kl} - u_k u_mE_{il} - u_iu_lE_{km} + u_k u_l E_{im}) ] \nonumber \\
& \quad +  [ u^p\nabla_p ( g_{im} E_{kl}  - g_{km} E_{il} - g_{il} C_{km} + g_{kl} E_{im}  ) \nonumber \\
& \qquad\qquad +2\varphi (g_{im}  E_{kl}  - g_{km} E_{il} - g_{il} E_{km} + g_{kl}  E_{im}  )] \nonumber
\end{align}
\begin{proof} 
Contraction of \eqref{Adati} with $u^j$ is:
\begin{align*}
u^j \nabla_i C_{jklm} + u^j\nabla_j C_{kilm} + u^j \nabla_k C_{ijlm}
= \tfrac{1}{n-3} (u_m \nabla_p C_{kil}{}^p + u_l \nabla_p C_{ikm}{}^p) \\
+ \tfrac{1}{n-3} \nabla_p [u^j ( g_{km} C_{ijl}{}^p 
+ g_{im}  C_{jkl}{}^p
+ g_{kl}  C_{jim}{}^p + g_{il} C_{kjm}{}^p )] \\
-\tfrac{1}{n-3}\varphi  u_p u^j
(g_{km} C_{ijl}{}^p 
+ g_{im}  C_{jkl}{}^p
+ g_{kl}  C_{jim}{}^p + g_{il} C_{kjm}{}^p )
\end{align*}
Where possible, the vector $u^k$ is taken inside covariant derivatives to take advantage of property \eqref{CCC}
\begin{align*}
& \nabla_i (u^jC_{jklm})-\varphi h_i^j C_{jklm} + u^j\nabla_j C_{kilm} + \nabla_k (u^jC_{ijlm})-\varphi h^j_k C_{ijlm}\\
& = \tfrac{1}{n-3} (u_m \nabla_p C_{kil}{}^p + u_l \nabla_p C_{ikm}{}^p) + \tfrac{1}{n-3} \nabla^p [ g_{km}(u_p E_{li}-u_lE_{pi}) \\
&+ g_{im} (u_lE_{pk} - u_p E_{lk}  )
+ g_{kl} (u_m E_{pi} -u_p E_{mi}) + g_{il} (u_p C_{mk} -u_m E_{pk} ] \\
& +\tfrac{1}{n-3}\varphi 
[g_{km} E_{il} 
- g_{im}  E_{kl}
- g_{kl}  E_{im} + g_{il} E_{km} ]
\end{align*}
\begin{align*}
\nabla_i (u_l E_{mk} -u_m E_{lk})-\varphi C_{iklm} - \varphi u_i (u_l E_{mk} -u_m E_{lk})+ u^j\nabla_j C_{kilm} \\
+ \nabla_k ( u_m E_{li}-u_l {\sf}C_{mi})  -\varphi C_{iklm}-\varphi u_k ( u_m E_{li}-u_l {\sf}C_{mi})\\
= \tfrac{1}{n-3} (u_m \nabla_p C_{kil}{}^p + u_l \nabla_p C_{ikm}{}^p) \\
+ \tfrac{1}{n-3} u^p\nabla_p [ g_{km} E_{li} 
- g_{im} E_{lk}  - g_{kl} E_{mi} + g_{il} C_{mk}  ] \\
+ \tfrac{1}{n-3} \nabla^p [- g_{km} u_lE_{pi} 
+ g_{im} u_lE_{pk} + g_{kl} u_m E_{pi}  - g_{il} u_m E_{pk}] \\
+\tfrac{n}{n-3}\varphi 
[g_{km} E_{il} 
- g_{im}  E_{kl}
- g_{kl}  E_{im} + g_{il} E_{km} ]
\end{align*}
\begin{align*}
(n-3)[u_l (\nabla_i  E_{mk} - \nabla_k E_{mi})
-u_m (\nabla_i E_{lk} - \nabla_kE_{li}) 
 -2\varphi C_{iklm} + u^j\nabla_j C_{kilm} ]\\
=  (u_m \nabla_p C_{kil}{}^p + u_l \nabla_p C_{ikm}{}^p) 
+  u^p\nabla_p [ g_{km} E_{li} 
- g_{im} E_{lk}  - g_{kl} E_{mi} + g_{il} C_{mk}  ] \\
- g_{km} u_l \nabla^pE_{pi} 
+ g_{im} u_l\nabla^pE_{pk} + g_{kl} u_m \nabla^pE_{pi}  - g_{il} u_m \nabla^pE_{pk} \\
+2\varphi 
[g_{km} E_{il} 
- g_{im}  E_{kl}
- g_{kl}  E_{im} + g_{il} E_{km} ]
\end{align*}
Contraction with $u^l$ yields the first result, \eqref{2divCtwisted}:
\begin{align*}
 \nabla_p C_{ikm}{}^p=(n-3)(\nabla_i  E_{km} - \nabla_k E_{im})\\
 + (n-2) [ u^p\nabla_p (u_i E_{km}-u_k E_{im} ) +2 \varphi (u_iE_{km} - u_k E_{im})]\\
+ (2u_k u_m +g_{km}) \nabla_p E_i{}^p - (2u_i u_m +g_{im}) \nabla_p E_k{}^p
\end{align*}
which is used to replace the covariant divergences  $\nabla_p C_{jkl}{}^p$ in the previous expression
\begin{align*}
(n-3)[u_l (\nabla_i  E_{mk} - \nabla_k E_{mi})
-u_m (\nabla_i E_{lk} - \nabla_kE_{li}) 
 -2\varphi C_{iklm} + u^j\nabla_j C_{kilm} ]\\
=  -u_m \{(n-3)(\nabla_i  E_{kl} - \nabla_k E_{il})
 + (n-2) [ u^p\nabla_p (u_i E_{kl}-u_k E_{il} ) +2 \varphi (u_iE_{kl} - u_k E_{il})]\\
+ (2u_k u_l +g_{kl}) \nabla_p E_i{}^p - (2u_i u_l +g_{il}) \nabla_p E_k{}^p\}\\
+ u_l \{(n-3)(\nabla_i  E_{km} - \nabla_k E_{im})
 + (n-2) [ u^p\nabla_p (u_i E_{km}-u_k E_{im} ) +2 \varphi (u_iE_{km} - u_k E_{im})]\\
+ (2u_k u_m +g_{km}) \nabla_p E_i{}^p - (2u_i u_m +g_{im}) \nabla_p E_k{}^p \}\\
+  u^p\nabla_p [ g_{km} E_{li} 
- g_{im} E_{lk}  - g_{kl} E_{mi} + g_{il} C_{mk}]  \\
- g_{km} u_l \nabla^pE_{pi} 
+ g_{im} u_l\nabla^pE_{pk} + g_{kl} u_m \nabla^pE_{pi}  - g_{il} u_m \nabla^pE_{pk} \\
+2\varphi 
[g_{km} E_{il} 
- g_{im}  E_{kl}
- g_{kl}  E_{im} + g_{il} E_{km} ]
\end{align*}
Some derivatives cancel, and we are left with 
\begin{align*}
(n-3)[ -2\varphi C_{iklm} - u^p\nabla_p C_{iklm} ]\\
=  -u_m \{ (n-2) [ u^p\nabla_p (u_i E_{kl}-u_k E_{il} ) +2 \varphi (u_iE_{kl} - u_k E_{il})]\}\\
+ u_l \{ (n-2) [ u^p\nabla_p (u_i E_{km}-u_k E_{im} ) +2 \varphi (u_iE_{km} - u_k E_{im})] \}\\
+  u^p\nabla_p [ g_{km} E_{li} 
- g_{im} E_{lk}  - g_{kl} E_{mi} + g_{il} C_{mk} ]  \\
+2\varphi 
[g_{km} E_{il} 
- g_{im}  E_{kl}
- g_{kl}  E_{im} + g_{il} E_{km} ]
\end{align*}
The final equation is obtained.
\end{proof}
\end{prop}

\end{document}